\documentclass[twocolumn]{aastex631}

\usepackage{multirow,array,amsmath,amssymb}
\usepackage[utf8]{inputenc}
\usepackage{CJK}
\usepackage{natbib} 
\usepackage{cleveref} 
\usepackage{booktabs}
\def\apjl{ApJL}
\def\apjs{ApJS}

\shorttitle{Calbration HIIGx HD with ANN}
\shortauthors{Zhang et al.}
\graphicspath{{./}{figures/}}

\begin{document}
\begin{CJK*}{UTF8}{gbsn}

\title{A Reliable Calibration of HII Galaxies Hubble Diagram with Cosmic Chronometers and Artificial Neural Network}

\author[0000-0002-1428-8311]{Jian-Chen Zhang(张建臣)}
\affiliation{Institute for Frontiers in Astronomy and Astrophysics of Beijing Normal University, Beijing 100875, China; tjzhang@bnu.edu.cn}
\affiliation{Department of Astronomy, Beijing Normal University, Beijing 100875, China}
\affiliation{College of Computer and Information Engineering, Dezhou University, Dezhou 253023, China}
\affiliation{Institute for Astronomical Science, Dezhou University, Dezhou 253023, China}

\author[0000-0003-0167-9345]{Kang Jiao(焦康)}
\affiliation{Institute for Frontiers in Astronomy and Astrophysics of Beijing Normal University, Beijing 100875, China; tjzhang@bnu.edu.cn}
\affiliation{Department of Astronomy, Beijing Normal University, Beijing 100875, China}
\affiliation{Dipartimento di Fisica e Astronomia ``Augusto Righi'', Alma Mater Studiorum Universit\`{a} di Bologna, via Piero Gobetti 93/2, I-40129 Bologna, Italy}

\author{Tingting Zhang}
\affiliation{College of Command and Control Engineering, PLA Army Engineering University, Nanjing 210000, China; 101101964@seu.edu.cn}

\author[0000-0002-3363-9965]{Tong-Jie Zhang(张同杰)}
\affiliation{Institute for Frontiers in Astronomy and Astrophysics of Beijing Normal University, Beijing 100875, China; tjzhang@bnu.edu.cn}
\affiliation{Department of Astronomy, Beijing Normal University, Beijing 100875, China}
\affiliation{Institute for Astronomical Science, Dezhou University, Dezhou 253023, China}

\author[0000-0002-9849-1762]{Bo Yu(于波)}
\affiliation{Institute for Astronomical Science, Dezhou University, Dezhou 253023, China}
\affiliation{School of Mathematics and Big Data, Dezhou University, Dezhou 253023, China}
\affiliation{Shandong Provincial Key Laboratory of Biophysics, Institute of Biophysics, Dezhou University, Dezhou 253023, China}

\begin{abstract}

The $L-\sigma$ relation of HII galaxies (HIIGx) calibrated by a distance indicator is a reliable standard candle for measuring the Hubble constant $H_0$. The most straightforward calibration technique anchors them with the first tier of distance ladders from the same galaxies. Recently another promising method that uses the cosmological model-independent Cosmic Chronometers (CC) as a calibrator has been proposed. We promote this technique by removing the assumptions about the cosmic flatness and using a non-parametric Artificial Neural Network for the data reconstruction process. We observe a correlation between the cosmic curvature density parameter and the slope of the  $L-\sigma$ relation, thereby improving the reliability of the calibration. Using the calibrated HIIGx Hubble diagram, we obtain a Type Ia Supernovae Hubble diagram free of the conventional assumption about $H_0$. Finally we get a value of $H_0=65.9_{-2.9}^{+3.0}~\mathrm{km~s^{-1}~Mpc^{-1}}$, which is compatible with latest Planck18 measurement.
\end{abstract}


\section{Introduction}
\label{sec:intro}

Hubble diagrams contain a wealth of information on the mass-energy content and evolutionary history of the Universe via the distance redshift relation they present. A Hubble diagram can be constructed using any astronomical source with standardizable intrinsic properties that are calibrated using known credible distance indicators. For example, Type Ia Supernovae (SNe Ia), which share the same explosion mechanism, thus have an intrinsic luminosity that can be used to measure the luminosity distance \citep{Riess1998,Perlmutter1999}. The $E_{p,i}-E_{iso}$ correlation of Gamma-ray bursts (GRBs) has no significant redshift dependence and thus can be used to measure distance \citep{Demianski2017}. The non-linear relationship between X-ray and UV emission in quasars enables a novel, highly accurate method for estimating the absolute luminosity, converting it to standard candles \citep{Lusso2017,Salvestrini2019,Lusso2020}. Fast radio bursts (FRBs) dispersion is a convolution of the cosmic distance element and the electron density, which enables the use of these events as cosmological distance measurements \citep{Kumar2019}. 

HII galaxies (HIIGx) are massive and compact aggregates of star formation. The starburst nearly entirely dominates the total luminosity of an HIIGx. The strong correlation between the luminosity $L(H\beta)$ in $H\beta$ lines and the ionized gas velocity dispersion $\sigma_v$ \citep{Terlevich:1981,Chavez:2014,Melnick2017} provides a tool for constructing the Hubble diagram and is then used to estimate $H_0$ after calibration with the first tier of distance ladders \citep{Chavez:2012,Arenas:2018}. Apart from this type of local calibration, a new method was recently presented that extends to high redshifts by reconstructing Cosmic Chronometers (CC) to obtain distance information \citep{Wu:2020}. They applied the prevalent Gaussian Process (GP) \citep {Seikel:2012} reconstruction in their approach and obtained an accurate calibration of the HIIGx Hubble diagram. However, their result is based on the assumptions that the Universe is flat and the observational data are correlated by a kernel function. Removing the assumptions will further improve the reliability of this calibration method, especially for the concern that the last assumption is relatively strong.

Artificial Neural Network (ANN) is a machine learning technique that excels at regression. It has recently been broadly utilized in astronomical studies. Such as distinguishing standard and modified gravity cosmologies \citep{Peel2019}, estimating cosmological parameters \citep{Wang:2020j}, simulating Cosmic Microwave background anisotropy maps\citep{Mishra2019}, creating weak lensing convergence map \citep{Mustafa2019}, discriminating cosmological models \citep{Schmelzle2017}, likelihood-free cosmological constraints \citep{WangYC2021} Etc. \cite{Wang:2020a} make use of  ANN  in reconstructing functions of the $H(z)$ and the $D_L(z)$ using OHD and SNe Ia data. They prove its reliability and superiority in characterizing data uncertainties.

In this article, we propose using the ANN and CC to obtain a reliable calibration of the HIIGx Hubble diagram without any assumptions. In section \ref{sec:calib}, we illustrate the principle of calibrating the HIIGx Hubble diagram. In section \ref{sec:measurement}, we describe the principle and methodology in the local measurement of $H_0$, specifically about using the calibrated HIIGx Hubble diagram in section \ref{sec:HII measure} and the cross-calibrated SNe Hubble diagram in section \ref{sec:SNe measure}. In section \ref{sec:results}, we present the data used in this article and present our results. In section \ref{sec:conc}, we draw our conclusions and engage in some discussion.

\section{Calibrating the HIIGx Hubble Diagram}
\label{sec:calib}

The luminosity of Balmer $\mathrm{H}\beta$ lines that emitted in HIIGx is strongly correlated with the ionized gas velocity dispersion $\sigma_v$ \citep{Terlevich:1981}, since both the intensity of ionizing radiation and the value of $\sigma_v$ grow as the starbust mass increases \citep{Siegel:2005}. This correlation can be approximated by the expression in \cite{Chavez:2012},
\begin{equation}
\lg \left[ \frac{L (\mathrm{H} \beta)}{\mathrm{erg}\,\mathrm{s}^{-1}} \right] = \alpha\, \lg \left[ \frac{\sigma_v (\mathrm{H} \beta)}{\mathrm{km} \, \mathrm{s}^{-1}} \right] + \kappa \ , \label{equ:Lsigmarel}
\end{equation}
where $\alpha$ and $\kappa$ are constant coefficients in this linear relation. Due to the low dispersion of this relation observed between $L(\mathrm{H}\beta)$ and $\sigma_v$, these galaxies and local HII regions can be utilized as standard candles \citep{Terlevich:2015,Wei:2016,Yennapureddy:2017,Leaf:2018}.   

The selection criteria ensure that the selected HIIGx are comprised of systems in which the luminosity is dominated by single and very young starbursts (less than $5 \,\mathrm{Myr}$ in age) \citep{Terlevich:2015}. Accordingly, the bolometric flux of the HIIGx can thus be regarded as primarily constituting the $\mathrm{H} \beta$ line. Therefore the luminosity distance-$D_L(z)$ of an HIIGx can be approximated using the luminosity $L (\mathrm{H} \beta)$ and
flux $F(\mathrm{H} \beta)$ associated to the $\mathrm{H} \beta$ line,
\begin{equation}
D_L = \left[ \frac{L (\mathrm{H} \beta)}{4 \pi F (\mathrm{H} \beta)} \right]^{1/2}\;,\label{eq:DL_HII}
\end{equation}
where $F (\mathrm{H} \beta)$ denotes the reddening corrected $\mathrm{H} \beta$ flux. By the  definition of distance modulus
\begin{equation}
\mu = 5 \lg D_L+ 25 \ ,
\label{eq:DLdef}
\end{equation} we can obtain 
\begin{equation}
\mu_{\mathrm{HII}} = 2.5[\alpha \lg \sigma_{v}(\mathrm{H} \beta) + \kappa- \lg F(\mathrm{H} \beta)] - 100.2,
\label{eq:muHII}
\end{equation}
while the error propagation is
\begin{equation}
\sigma^2_{\mu_\mathrm{HII}}(\alpha,z_i)=2.5^2[(\alpha\sigma_{\lg\sigma_v})^2+\sigma^2_{\lg F}],
\label{eq:sigmamuhii}
\end{equation}
where $\sigma_{\lg\sigma_v}$ and $\sigma_{\lg F}$ are the measurement error ${\lg\sigma_v}$ and $\lg F$ respectively.

If we can measure the $D_L$ at the same redshift as the observed HIIGx, we are able to calibrate the slop ($\alpha$) and intercept ($\kappa$) of the $L-\sigma$ relation. Locally, it is not difficult to anchor the HIIGx with reliable distance modulus determined by other distance indicators within the same galaxy, such as the period-luminosity relation of Cepheid. For a higher redshift range that the first tier of the distance ladder is not applicable, we might obtain the anchors by integrating the inverse of the Observation $H(z)$ Data (OHD, also known as CC) across redshift,
\begin{equation}
    \frac{D_{L}}{(1+z)}=\left\{\begin{array}{ll}
\frac{D_{H}}{\sqrt{\Omega_{k0}}} \sinh \left[\sqrt{\Omega_{k0}} D_{C} / D_{H}\right] & \Omega_{k0}>0 \\
D_{C} & \Omega_{k0}=0 \\
\frac{D_{H}}{\sqrt{\left|\Omega_{k0}\right|}} \sin \left[\sqrt{\left|\Omega_{k0}\right|} D_{C} / D_{H}\right] & \Omega_{k0}<0
\end{array}\right.
\label{eq:DL}
\end{equation}
where $\Omega_{k0}$ is the curvature energy desity parameter, the Hubble radius $D_{H}=c/H_0$, and the comoving distance 
\begin{equation}
    D_{C}=c \int_{0}^{z} \frac{d z^{\prime}}{H\left(z^{\prime}\right)},
\label{eq:DC}
\end{equation}
$c$ denotes the speed of light, and $H(z^\prime)$ denotes the Hubble parameter at redshift $z^\prime$. If we want to get the $D_C$ at any redshift of interest, a reconstruction based on the OHD is a prerequisite. Reconstructing $H(z)$ would complicate the covariance propagation, therefore we choose the same method as \cite{Wu:2020} that reconstruct $c/H(z)$. We do the integration using
\begin{equation}
    D_{C}\simeq\sum \frac{c}{H(z_i)} \Delta z_{i},
\end{equation}
which strictly holds if the redshift interval $\Delta z_{i} = \frac{1}{2}(z_{i-1}-z_{i+1})$ is smaller enough.
The covariance of the reconstructed data will propagate to the $D_C$ in an approximate form of 
\begin{equation}
    \operatorname{Cov}_{D_{C}}(z_i,z_j)= \sum_{ l}^{i}\sum_{k}^{j}\operatorname{Cov}\big(\frac{c}{H_k},\frac{c}{H_l}\big)\Delta z_{k}\Delta z_{l},
\end{equation}
which further propagate to the covariance of $\mu$ as
\begin{equation}
  \operatorname{Cov}_{\mu}(H_0, \Omega_{k0};z_i,z_j) = \mathcal{C}_i\mathcal{C}_j\operatorname{Cov}_{D_{C}}(z_i,z_j),
  \label{eq:covmuohd}
\end{equation}
where the coefficient $\mathcal{C}_i$ in error propagation is a function of $\Omega_k$ and $H_0$,
\begin{equation}
\mathcal{C}_i
\frac{\ln10}{5}=\left\{\begin{array}{ll}
\frac{\sqrt{\Omega_{k0}}}{D_{H}}\coth \left[\sqrt{\Omega_{k0}} D_{Ci} / D_{H}\right] & \Omega_{k0}>0 \\
\frac{1}{D_C}& \Omega_{k0}=0 \\
 \frac{\sqrt{\left|\Omega_{k0}\right|}}{D_{H}}\cot\left[\sqrt{\left|\Omega_{k0}\right|} D_{Ci} / D_{H}\right]& \Omega_{k0}<0
\end{array}\right. .
\end{equation}

\begin{figure*}[tb!]
    \gridline{\includegraphics[width=.5\textwidth]{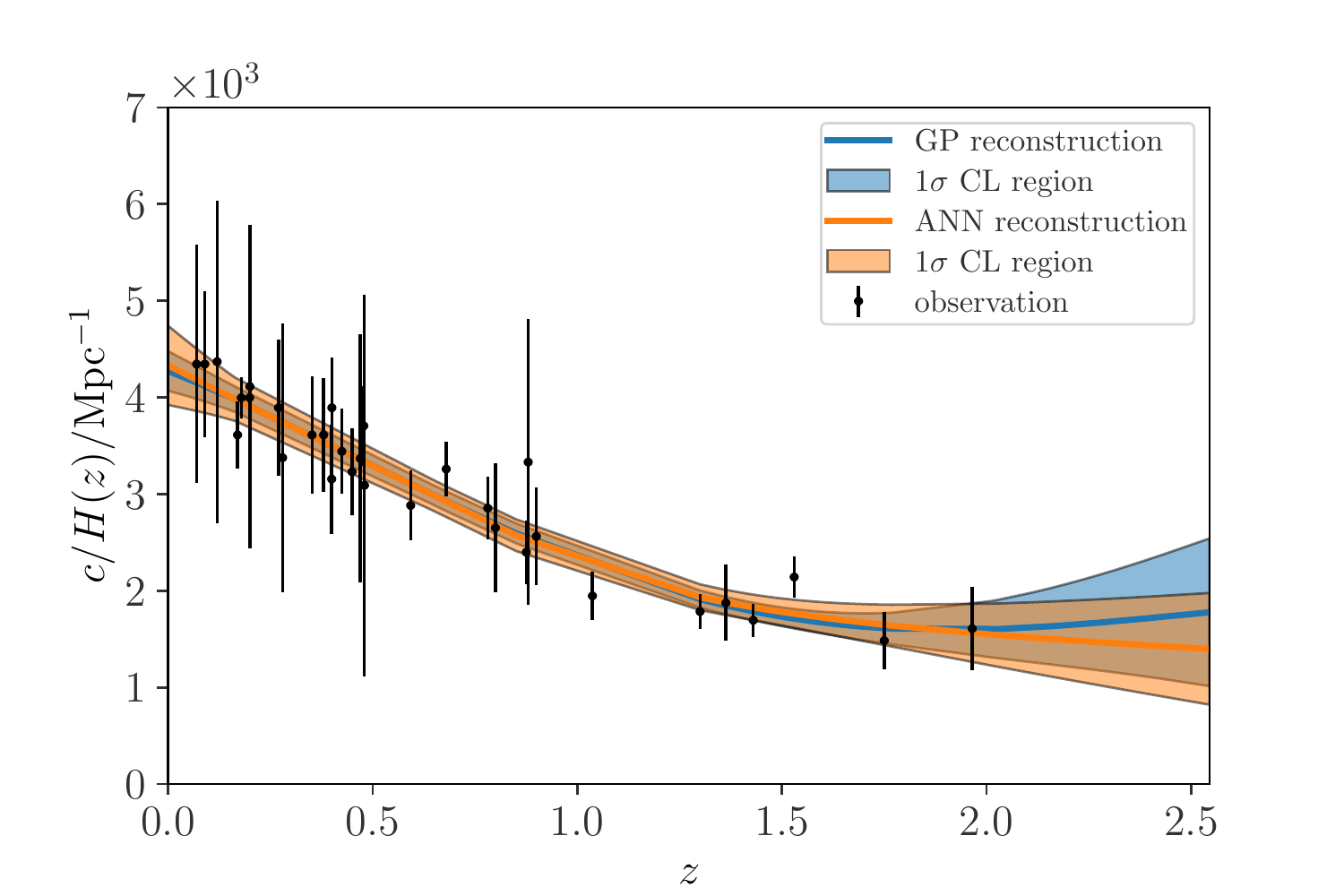}\includegraphics[width=.5\textwidth]{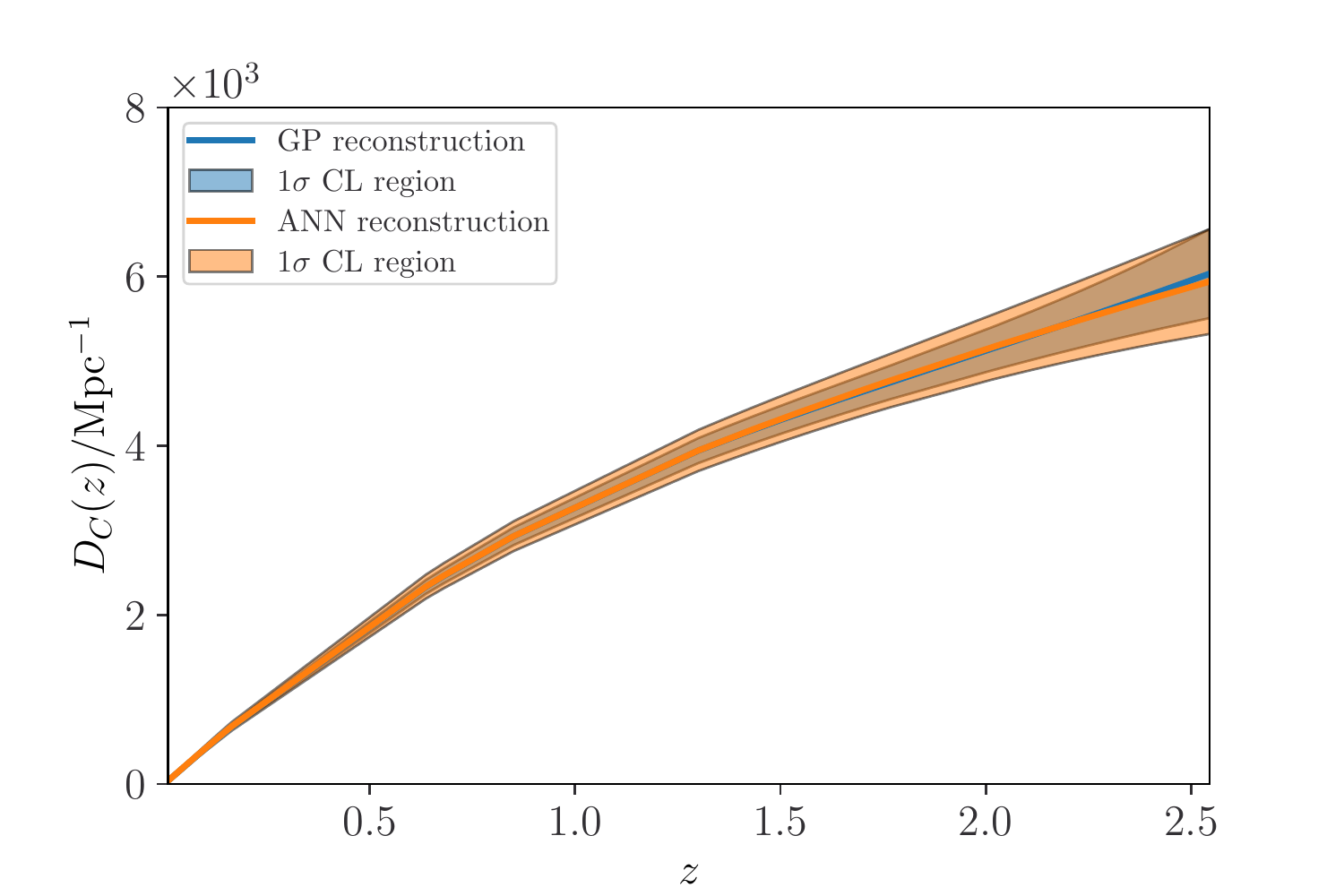}}
    \caption{Left: reconstructions of $c/H(z)$ using Gaussian Process (blue) and Artificial Neural Network (orange). The solid central lines represent the reconstructed values, whereas the shaded regions represent the 1 $\sigma$ confidence intervals. The black error bars are the 32 OHD that are used as the basis in our reconstructions. Right: the corresponding reconstructions of the comoving distances, while 1 $\sigma$ regions are calculated based on the upper and lower bounds of the $c/H(z)$ in the left panel. }
    \label{fig:rec}
\end{figure*}

The HIIGx Hubble diagram can be calibrated without any assumption of the cosmological model or the geometry of the Universe, using the Bayesian inference method by defining the log-likelihood function
\begin{equation}
    \begin{aligned}
\ln \mathcal{L}=
&-\frac{1}{2}\ln \Big((2 \pi)^{N}	\det\boldsymbol{\operatorname{Cov}}\Big)-\frac{1}{2}\Delta\boldsymbol{\mu}^{\top} \boldsymbol{\operatorname{Cov}}^{-1}\Delta\boldsymbol{\mu},
\end{aligned}
\label{def:lnL}
\end{equation}
where the distance modulus difference vector is 
\begin{equation}
    \Delta\boldsymbol{\mu}=\boldsymbol{\mu_\text{HII}}(\alpha,\kappa;\boldsymbol{z}_\text{HII})-\boldsymbol{\mu^\text{rec}_\text{OHD}}(H_0, \Omega_{k0};\boldsymbol{z}_\text{HII}),
\end{equation}
and the corresponding total covariance matrix is
\begin{equation}
    \boldsymbol{\operatorname{Cov}} = \boldsymbol{\operatorname{Cov}^\text{rec}_{\mu_\text{OHD}}}+\operatorname{diag}(\sigma^2_{\mu_\mathrm{HII}}),
\end{equation}
where $\boldsymbol{\operatorname{Cov}^\text{rec}_{\mu_\text{OHD}}}$ is covariance matrix of the reconstructed $\mu_\text{OHD}$ that described in ~\cref{eq:covmuohd}, while the covariance matrix of $\mu_\mathrm{HII}$ is diagonal that described in~\cref{eq:sigmamuhii}.

\section{Local Measurement the $H_0$}
\label{sec:measurement}
In cosmography, the $D_L(z)$, as a function of redshift, can be expanded into the Taylor series as
\begin{equation}
D_L(z) = \frac{cz}{H_0} \left[\mathcal{D}_L^0+\mathcal{D}_L^1z+\mathcal{D}_L^2z^2+\mathcal{O}(z^3)\right]\;,\label{eq:DL_expansion}
\end{equation}
where
\begin{equation}
\begin{aligned}
&\mathcal{D}_L^0=1\\
&\mathcal{D}_L^1=-\frac{1}{2}(1-q_0)\\
&\mathcal{D}_L^2=-\frac{1}{6}(1-q_0-3q_0^2+j_0)  \;.
\end{aligned}
\end{equation}
The $q_0$ and $j_0$ in above equation are the deceleration parameter the jerk parameter respectively, which are the dimensionless second and third derivative of the scale factor $a(t)$ with respect to cosmic time $t_0$. And the first-order approximation at low redshifts in ~\cref{eq:DL_expansion} is the famous Hubble's law,
\begin{equation}
D_L(z) = \frac{cz}{H_0}\ .
\label{eq:HubbleLaw}
\end{equation}
It is independent of the mass-energy content of the Universe but only valid locally ($z\lesssim0.1$).

\subsection{Measurement with the Calibrated HIIGx Hubble Diagram}
\label{sec:HII measure}
Once the Hubble diagram of HIIGx had been properly calibrated, it could be widely used to constrain cosmological model parameters. For instance, the $H_0$ could be constrained directly using only the local HIIGx measurements, or it could be constrained together with energy density parameters using the entire sample and assuming a cosmological model. We will concentrate on the direct measurement of $H_0$ in this article, which can be achieved by performing a Markov Chain Monte Carlo (MCMC) to maximize the log-likelihood function that described in ~\cref{def:lnL} with
\begin{equation}
    \Delta\boldsymbol{\mu}=\boldsymbol{\mu^\text{local}_\text{HII}}(\alpha,\kappa;\boldsymbol{z}^\text{local}_\text{HII})-\boldsymbol{\mu_\text{th}}(H_0;\boldsymbol{z}^\text{local}_\text{HII}),
\end{equation}
where the theoretical $H_0$-dependent distance module $\boldsymbol{\mu_\text{th}}(H_0,\boldsymbol{z}^\text{local}_\text{HII})$ is determined by substituting ~\cref{eq:HubbleLaw} into ~\cref{eq:DLdef} at local redshifts ($z<0.1$) of HIIGx, and the corresponding covariance matrix is sub-matrix of the full $\boldsymbol{\operatorname{Cov}^\text{rec}_{\mu_\text{OHD}}}$ ,
\begin{equation}
    \boldsymbol{\operatorname{Cov}} = \boldsymbol{\operatorname{Cov}^\text{rec}_{\mu_\text{OHD}}}(\boldsymbol{z}^\text{local}_\text{HII},\boldsymbol{z}^\text{local}_\text{HII}).
\end{equation}
To achieve a better constraint on the $H_0$ and the calibration of the HIIGx Hubble diagram, we make a joint constraint by maximizing the summation of the log-likelihood function
\begin{equation}
\begin{aligned}
    \ln \mathcal{L} = &\ln \mathcal{L}(\boldsymbol{\mu_\text{HII}},\boldsymbol{\mu^\text{rec}_\text{OHD}}|H_0, \Omega_{k0}, \alpha, \kappa) \\&+  \ln \mathcal{L}(\boldsymbol{\mu^\text{local}_\text{HII}}|H_0, \alpha, \kappa)   
\end{aligned}
\end{equation}
which are described in section~\ref{sec:calib} and here above.

\begin{figure*}[tb!]
    \centering
	\includegraphics[width=.7
	\textwidth]{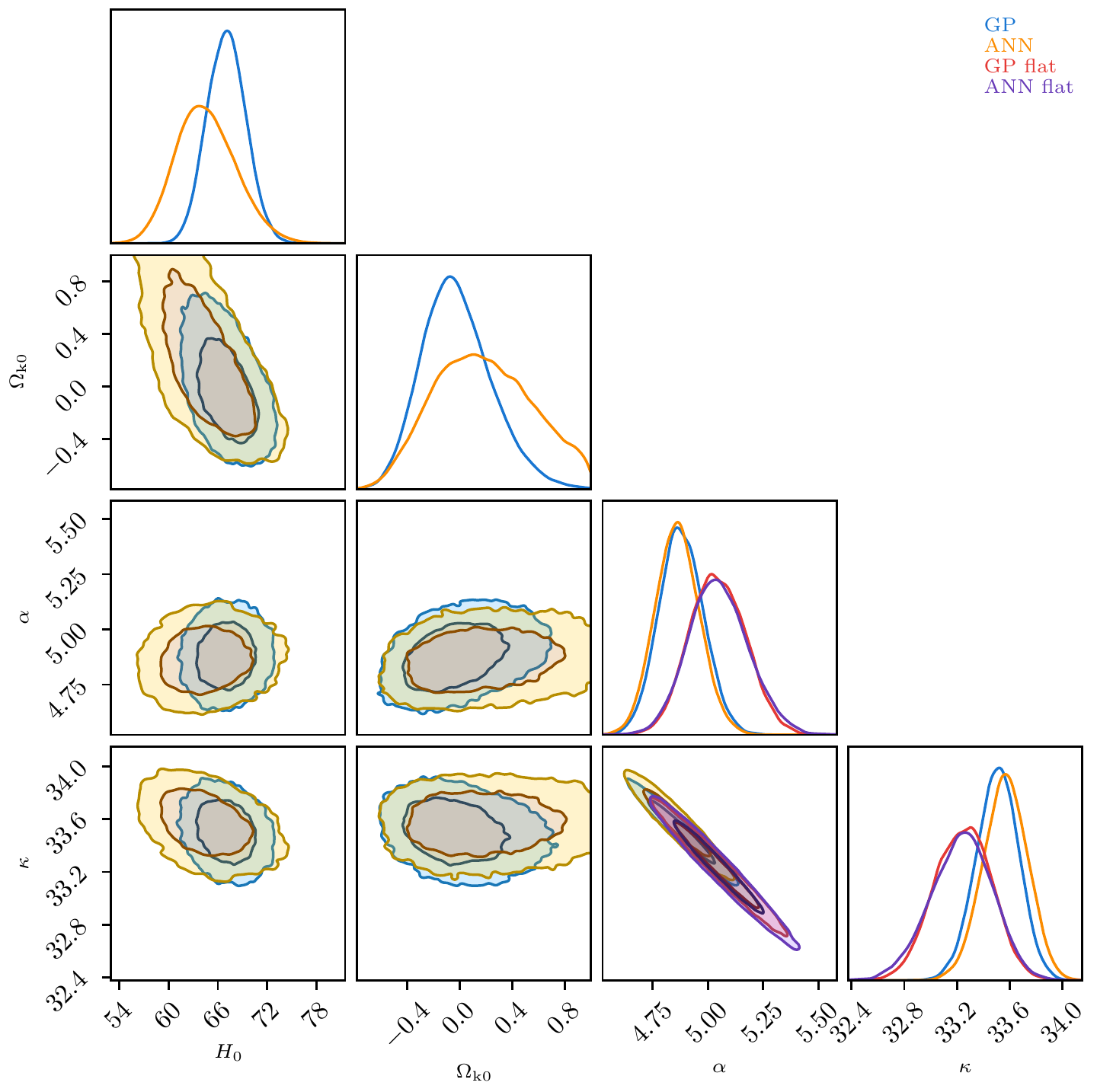}
	\caption{68\% and 95\% confidence contours for the parameters involved in calibrating the HIIGx Hubble diagram with OHD. Results from the fit of not assuming flat Universe are show in blue and orange contours for using the GP and ANN reconstructed $\mu_\mathrm{OHD}$ data respectively, whereas the red and violet contours are that from the assuming a flat Universe. }
	\label{fig:ohd+hii}
\end{figure*}

\subsection{Measurement with the Cross-Calibrated SNe Hubble Diagram}
\label{sec:SNe measure}
SNe Ia is widely used as secondary standard candles to measure luminosity distances $D_L$ because the peak luminosities of light curves of all SNe Ia are nearly identical. By identifying an SNe Ia that shares a host galaxy with a Cepheid variable, one can determine the host galaxy's $D_L$ via its period-luminosity relation. Principally, by combining the $D_L$ with the peak apparent magnitude $m_{\max}$ of the SNe Ia, we can obtain the peak absolute magnitude $M_{\max}$ of all the  SNe Ia using
\begin{equation}
\mu = 5 \lg \frac{D_L}{\mathrm{Mpc}} + 25 = m_{\max}-M_{\max}.
\end{equation} Then the $D_L$ of an arbitrary SNe Ia can be easily obtained by measuring its $m_\text{max}$.  

However, the $M_{\max}$ of SNe Ia are not identical but also related to the shapes can colors of the light curves. Taking this into account, the formula for SNe Ia distance modulus should be modified by including perturbations of shapes $x$ and color $c$ as shown in \cite{Guy:2007},
\begin{equation}
\mu_{B}(\alpha_X,\beta,M_{B})=m_{B}^{\rm max}-M_{B}+\alpha_X X-\beta C,
\label{eq:mu_B}
\end{equation}
where the subscript $B$ stands for B band, while $\alpha$ and $\beta$ are nuisance parameters for modification. 
It is well known that the $M_B$ is strongly correlated with $H_0$. As a convention in SNe Ia cosmology, it is safe to constrain parameters that are not very much dependent on $M_B$, based on an assumed value of $H_0 = 70\ \mathrm{km\ s^{-1}\ Mpc^{-1}}$. However, if the objective is to precisely measure the $H_0$, we must first obtain an accurate measurement of the $M_B$, which should also be modified according to the stellar mass ($M_\mathrm{stellar}$), in a step function form of 
\begin{equation}
    M_{B}=\left\{\begin{array}{ll}
M_{B}^{1} & \text { if }\  M_{\text {stellar }}<10^{10} M_{\odot} \\
M_{B}^{1}+\Delta_{M} & \text { otherwise. }
\end{array}\right. ,
\end{equation}
where the modified term $\Delta_{M}$ only takes effect for massive objects.
The \cref{eq:mu_B} can be written in matrix form of
\begin{equation}
    \boldsymbol{\mu}_{B}=\mathbf{A} \boldsymbol{\eta}-\boldsymbol{M}_{B},
\end{equation}
where the light-curve parameters vector is 
\begin{equation}
    \boldsymbol{\eta}= \left(m_{B1},X_{1},C_{1}, \cdots,m_{Bn},X_{n},C_{n}\right),
\end{equation}
and the coefficient matrix is 
\begin{equation}
    \mathrm{A}=\mathrm{A}_{0}+\alpha \mathrm{A}_{1}-\beta \mathrm{A}_{2},
\end{equation}
whose elements can be decide according to 
\begin{equation}
    \left(\mathbf{A}_{k}\right)_{i, j}=\delta_{3 i, j+k}.
\end{equation}
The covariance matrix of the $\eta$ consists of statistical uncertainties of light-curve fitting and seven sources of systematic uncertainties, including the covariance of calibration, light curve model, bias correction,  mass step uncertainty, Milky Way dust extinction correction, peculiar velocity corrections, contamination of Hubble diagram by non-Ia. For detail descriptions please referred to \cite{Betoule:2014} .
The covariance further propagate to $\boldsymbol{\mu_B}$ in the form of
\begin{equation}
    \begin{aligned}
        \operatorname{Cov}_{\boldsymbol{\mu_B}}=\mathbf{A}\operatorname{Cov}_{\eta} \mathbf{A}^{\top}&+\operatorname{diag}\left(\frac{5 \sigma_{z}}{z \ln 10}\right)^{2}\\&+\operatorname{diag}\left(\sigma_{\text {lens }}^{2}\right)+\operatorname{diag}\left(\sigma_{\mathrm{coh}}^{2}\right),
    \end{aligned}
\end{equation}
taking account of the uncertainty in cosmological redshift due to peculiar velocities ($\sigma_{z}$), the variation of magnitudes caused by gravitational lensing ($\sigma_{\text {lens }}$), and the intrinsic variation in SN magnitude not described by the other terms ($\sigma_{\mathrm{coh}}$).

Once we get the HIIGx Hubble diagram which is precisely calibrated using the method described in section~\ref{sec:calib}, we can take distance anchors from it to cross calibrate the SNe Ia Hubble diagram. We define the log-likelihood function in~\cref{def:lnL} with the distance modulus difference vector in the form of
\begin{equation}
    \Delta \boldsymbol{\mu}=\boldsymbol{\mu_B}\left(\alpha_x,\beta, M_B^1, \Delta M;\boldsymbol{z}_\text{SNe}\right)-{\boldsymbol{\mu}^\text{rec}_\mathrm{HII}}(\boldsymbol{z}_\text{SNe}),
    \label{eq:deltamu3}
\end{equation}
where ${\boldsymbol{\mu}^\text{rec}_\mathrm{HII}}(\boldsymbol{z}_\text{SNe})$ is reconstructed based on the calibrated $\boldsymbol{\mu}^\text{calib}_{\mathrm{HII}}$ described in section~\ref{sec:calib},
whose error propagation is given by
\begin{gather}
    \begin{aligned}
    \sigma^2_{\boldsymbol{\mu}^\text{calib}_{\mathrm{HII}}}= 2.5^2[\alpha^2\sigma^2_{\lg\sigma}+\sigma_{\alpha}^2\lg^2\sigma+\sigma^2_\kappa +\sigma^2_{\lg F}],
    \label{eq:sigma_mu_obs}
    \end{aligned}
\end{gather}
where $\sigma_{\alpha}$, $\sigma_\kappa$, are the calibration errors of $\alpha$ and $\kappa$, while $\sigma_{\lg\sigma_v}$, $\sigma_{\lg F}$ are measurement errors of $\lg\sigma(\rm H\beta)$ and $\lg F(\rm H\beta)$ respectively. The total covariance matrix corresponding to~\cref{eq:deltamu3} is 
\begin{equation}
    \boldsymbol{\operatorname{Cov}}=\boldsymbol{\operatorname{Cov}}_{\boldsymbol{\mu}_{B}}+\boldsymbol{\operatorname{Cov}}^\text{rec}_{\boldsymbol{\mu}_\mathrm{HII}}.
\end{equation}
We take the same procedure in section \ref{sec:HII measure} to determine the value of $H_0$ using this cross calibrated SNe Hubble diagram, therefore we define
\begin{equation}
    \Delta \boldsymbol{\mu}=\boldsymbol{\mu}^\text{local}_B\left(\alpha_x,\beta, M_B^1, \Delta M;\boldsymbol{z}^\text{local}_\text{SNe} \right)-\boldsymbol{\mu_\text{th}}(H_0;\boldsymbol{z}^\text{local}_\text{SNe})
\end{equation}
only using the local samples and get the covariance matrix
\begin{equation}
    \boldsymbol{\operatorname{Cov}} = \boldsymbol{\operatorname{Cov}^\text{rec}_{\mu_\text{HII}}}(\boldsymbol{z}^\text{local}_\text{SNe},\boldsymbol{z}^\text{local}_\text{SNe})
\end{equation}
correspondingly. We perform a MCMC to maximize the summation of the log-likelihood function
\begin{equation}
\begin{aligned}
    \ln \mathcal{L} = &\ln \mathcal{L}(\boldsymbol{\mu}_B,\boldsymbol{\mu^\text{rec}_\text{HII}}|\alpha_{x}, \beta, M_{B}^{1}, \Delta M) \\&+  \ln \mathcal{L}(\boldsymbol{\mu_\textit{B}^\text{local}}|H_0,\alpha_{x}, \beta, M_{B}^{1}, \Delta M)   
\end{aligned}
\end{equation}
and get the $H_0$ and calibrated SNe Hubble diagram simultaneously.

\begin{figure}[tb!]
    \centering
    \includegraphics[width=.5\textwidth]{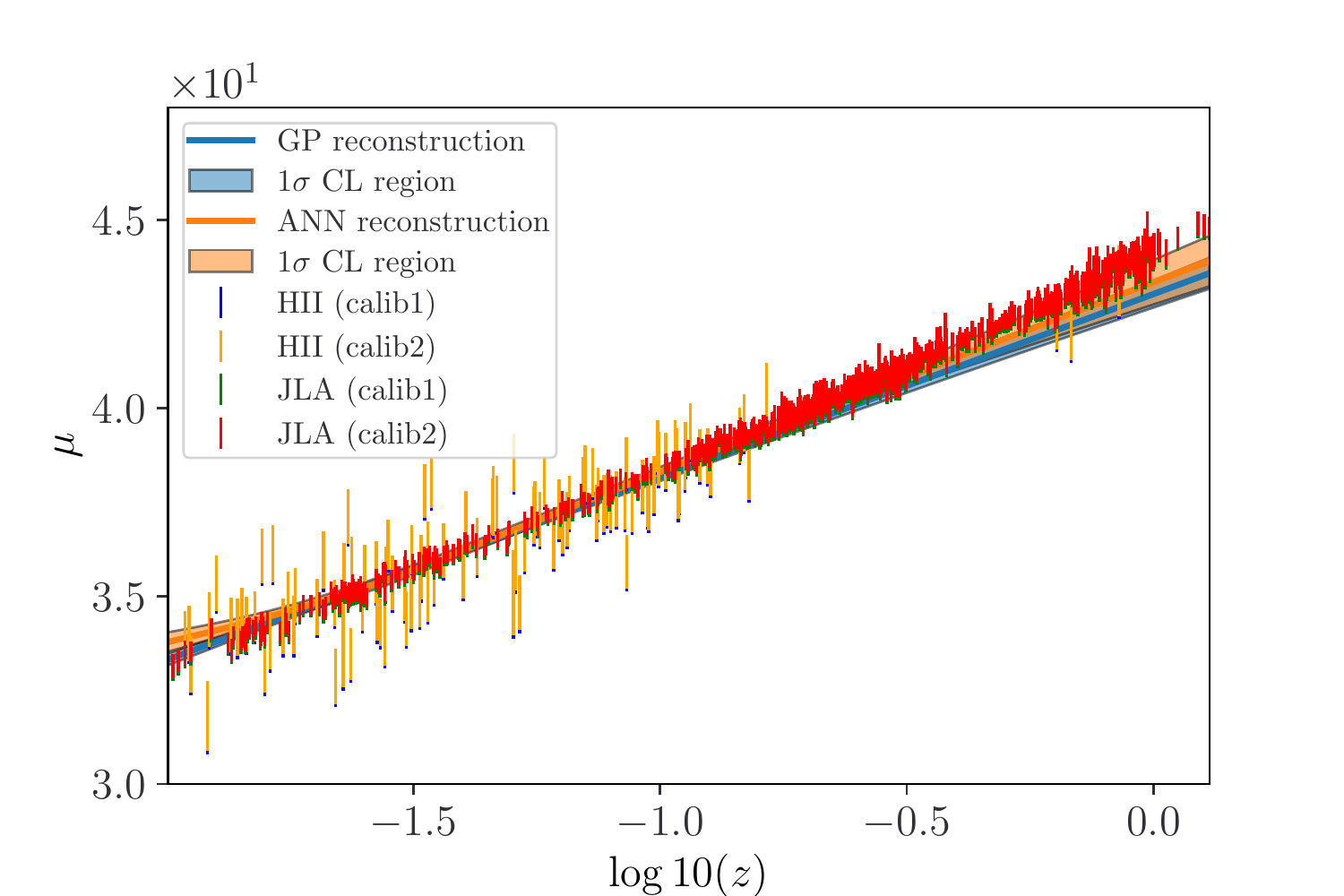}
    \caption{Hubble diagrams of HIIGx and JLA SNe Ia samples. Error-bars represent the calibrated distance modulus of the HIIGx and JLA sample and their corresponding 1 $\sigma$ uncertainties, labeled by the sample names and calibration methods, with calib1 and calib2 stand for using Gaussian Process (GP) and Artificial Neural Network (ANN) reconstructed data respectively. The blue (orange) solid curves and shaded regions are the GP (ANN) reconstructed $\mu^\text{rec}_\mathrm{HII}$ and their corresponding 1 $\sigma$ confidence regions at the redshift of JLA sample.}
    \label{fig:mu}
\end{figure}

\begin{figure*}[tb!]
    \centering
    \includegraphics[width=.9\textwidth]{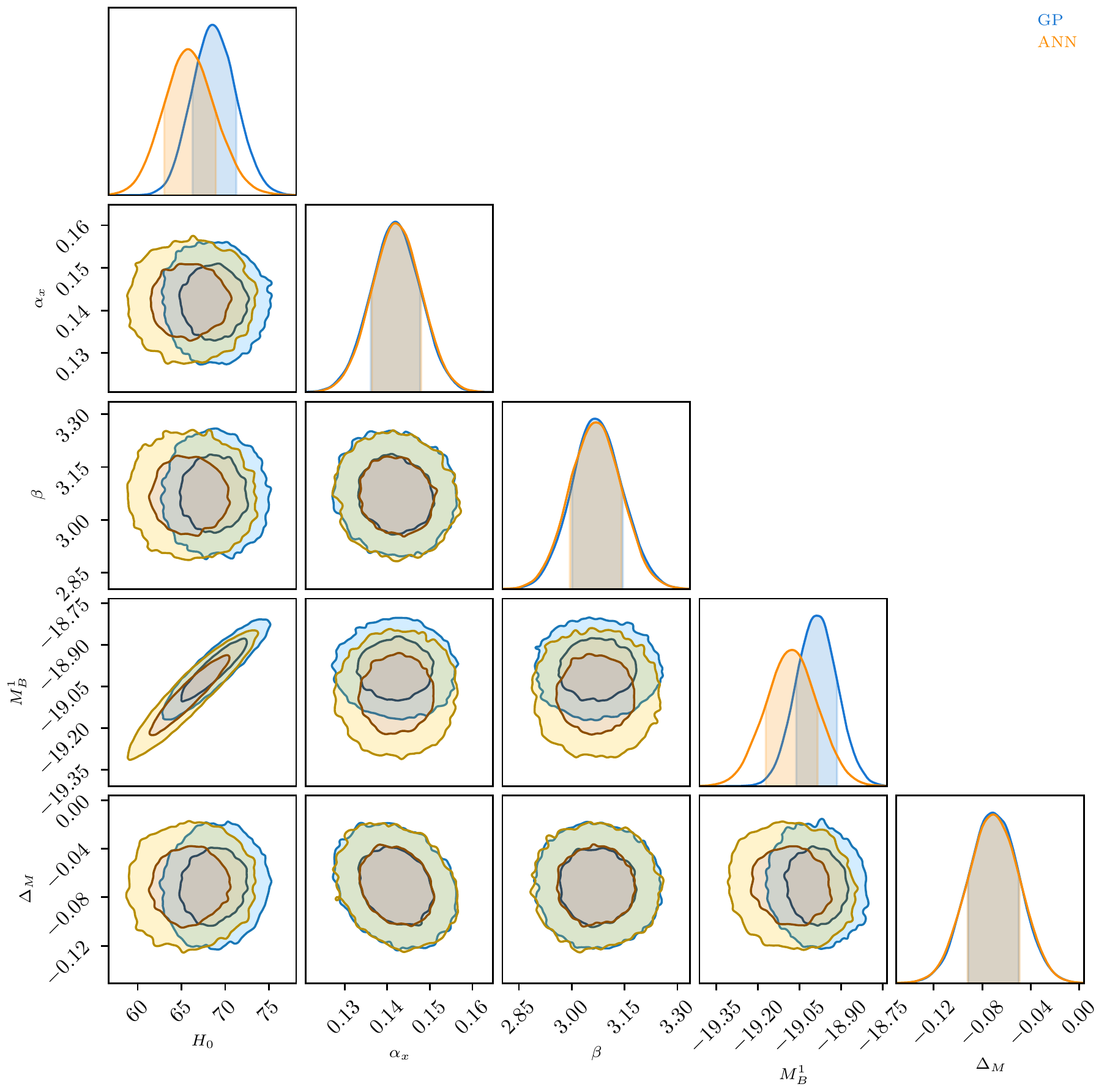}
    \caption{68\% and 95\% confidence contours for the parameters involved in calibrating the SNe Hubble diagram with calibrated HIIGx data. Blue (orange) contours result from the fit of using the GP (ANN) reconstructed $\mu_\mathrm{HII}$ data and the full JLA sample.}
    \label{fig:hii+sne}
\end{figure*}

\section{Data and Results}
\label{sec:results}
We applied the methods that are described in section \ref{sec:calib} and \ref{sec:measurement} to the following list of data in detail.

\begin{enumerate}
	\item[(1)]  \textit{OHD}. We take the updated full sample of cosmological model-independent OHD which are measured according the equation
	\begin{equation}
	H(z) = -\frac{1}{1+z}\frac{\mathrm{d}z}{\mathrm{d}t} \simeq -\frac{1}{1+z}\frac{\Delta z}{\Delta t}
	\end{equation}
	by measuring the time derivative of redshift $\mathrm{d}z / \mathrm{d}t$ without assuming any cosmological model. The derivative is measured by taking the differential age of two massive and passively evolving galaxies at different redshifts as an approximation. A total number of 32 OHD (listed in table~\ref{table:1}) are given by \cite{Jimenez:2003,Simon:2005,Stern:2010,Moresco:2012,Zhang:2014,Moresco:2015,Moresco:2016,Ratsimbazafy:2017,Borghi2022ApJb,Jiao2022}.  Considering the last two measurements are not fully independent because they are derived from the same sample of passive galaxies sample, and their covariance is not clear yet, we decide only use the latest \cite{Jiao2022} measurement taking advantage of $\sim1/\sqrt{2}$ fraction of systematic uncertainty that is improved by approximately double number galaxies were used. One could either use the \cite{Borghi2022ApJb} measurement or both of them if the covariance is cleverly considered.

	\item[(2)]  \textit{HIIGx}. We use the full sample of HIIGx composed by \cite{Gonzalez2021MNRAS}, which includes 107 local HII galaxies \citep{Chavez:2014} recalibrated by \citep{Gonzalez2019MNRAS} in the redshift range of $0.0088 \leq z \leq 0.16417$, and 74 high$\text{-}z$ HII galaxies in the redshift range of $0.63427 \leq z \leq 2.545$. The fluxes and gas velocity dispersions (along with their uncertainties) of HIIGx and GEHR that we use in this article are all referred to from this catalogue.

	\item[(3)] \textit{JLA}. This SNe Ia catalog is composed by \cite{Betoule:2014} from observations obtained by the SDSS-II and SNLS collaborations. The data set includes several low-redshift samples ($z < 0.1$), all three seasons from the SDSS-II ($0.05 < z < 0.4$), and three years from SNLS ($0.2 < z < 1$), and it totals 740 spectroscopically confirmed SNe Ia with high quality light curves. The light curve parameters $ \left(m_{B1},X_{1},C_{1}, \cdots,m_{Bn},X_{n},C_{n}\right)$ and the corresponding covariance matrix are all obtained from this catalogue. 
\end{enumerate}

\begin{table}
\begin{center}
\caption{All the Observational $H(z)$ Data derived from the Cosmic Chronomters (CC) method}
\begin{tabular}{ccccc}
\hline
\hline
 \emph{z} & $H(z)$ & $\sigma_{H}$ & References \\
\hline
0.09	&	$	69$&$12	$   &	\citet{Jimenez:2003} \\
\hline
0.17	&	$	83$&$8	$	 &	\\
0.27	&	$	77$&$14	$	 &	\\
0.4	&	$	95$&$17	$	 &	 \\
0.9	&	$	117$&$23	$	 &	\citet{Simon:2005} \\
1.3	&	$	168$&$17	$	 &	\\
1.43	&	$	177$&$18	$	 &	\\
1.53	&	$	140$&$14	$	 &	\\
1.75	&	$	202$&$40	$	 &	\\
\hline
0.48	&	$	97$&$62	$	 &	\citet{Stern:2010} \\
0.88	&	$	90$&$40	$	 &	\\
\hline
0.1791	&	$	75$&$4	$	 &	\\
0.1993	&	$	75$&$5	$	 &	\\
0.3519	&	$	83$&$14	$	 &	\\
0.5929	&	$	104$&$13	$	 &	\citet{Moresco:2012} \\
0.6797	&	$	92$&$8	$	 &	\\
0.7812	&	$	105$&$12	$	 &	\\
0.8754	&	$	125$&$17	$	 &	\\
1.037	&	$	154$&$20	$	 &	\\
\hline
0.07	&	$	69$&$19.6	$	 &	\\
0.12	&	$	68.6$&$26.2	$	 &	\citet{Zhang:2014} \\
0.2	&	$	72.9$&$29.6	$	 &	\\
0.28	&	$	88.8$&$36.6	$	 &	\\
\hline
1.363	&	$	160$&$33.6	$	 &	\citet{Moresco:2015} \\
1.965	&	$	186.5$&$50.4	$	 &	\\
\hline
0.3802	&	$	83$&$13.5	$	 &	\\
0.4004	&	$	77$&$10.2	$	 &	\\
0.4247	&	$	87.1$&$11.2	$	 &	\citet{Moresco:2016} \\
0.4497	&	$	92.8$&$12.9	$	 &	\\
0.4783	&	$	80.9$&$9	$	 &	\\
\hline
0.47   &      $      89$&$34        $       &   \citet{Ratsimbazafy:2017} \\
\midrule
0.75 & $98.8$ &$33.6$ & \citet{Borghi2022ApJb}\\
\hline
0.80 & $113.1$ &$28.5$ & \citet{Jiao2022}\\
\bottomrule
\end{tabular}\label{table:1}
\end{center}
{\raggedright * $H(z)$ and $\sigma_{H}$ are in the unit of $\rm km\ s^{-1}\ Mpc^{-1}$ \par}
\end{table}

For the calibration of the HIIGx Hubble diagram, we reconstruct $D_C$ at the redshifts of HIIGx sample based on the OHD, whereas for the calibration of the JLA SNe Ia Hubble diagram we reconstruct ${\boldsymbol{\mu}_\mathrm{HII}}$ at redshifts of JLA sample based on the pre-calibrated HIIGx Hubble diagram. We perform the following two methods to test the reliability of reconstructions.
GP reconstructs values that are connected in each pair of data points by assuming a covariance function $k(z, \tilde{z})$. We use \textsc{GaPP} \citep{Seikel:2012} and adopt the squared exponential kernal function
\begin{equation}
k(z, \tilde{z})=\sigma_{f}^{2} \exp \left(-\frac{(z-\tilde{z})^{2}}{2 \ell^{2}}\right),
\label{eq:kernal}
\end{equation}
where the signal variance $\sigma_{f}$ and length scale $\ell$ are two hyperparameters that should be optimized. The covariance matrix
\begin{equation}
    \boldsymbol{\operatorname{Cov}} = \boldsymbol{ K}_{**}-  \boldsymbol{ K}_{o*}^\top
   \boldsymbol{ K}_{oo}^{-1} \boldsymbol{K}_{o*},
\end{equation}
where the subscripts $\textit{o}$ and $*$ stand for calculating \cref{eq:kernal} at the redshifts of the observational data and the reconstructed data, with the first and second position stand for the rows and columns respectively.
ANN is an alternative non-parametrical reconstruction method free of assumptions that can improve reliability. We use \textsc{ReFANN} as our second tool, which is validated in reconstructing OHD and $D_L(z)$ \citep{Wang:2020j}. The covariance matrix of can be estimated by the method described in section 6.2 of \cite{Wang:2020j},
\begin{equation}
    \begin{array}{l}
\operatorname{Cov}\left(\frac{c}{H_i},\frac{c}{H_j}\right)=\frac{1}{N}\sum_{k=1}^{N}\left[\left(\frac{c}{H_i}_{k}-\bar{\frac{c}{H_i}}\right)\left(\frac{c}{H_j}_{k}-\bar{\frac{c}{H_j}}\right)\right],
\end{array}
\end{equation}
which averaged over a large number of reconstructions (i.e., $N=1000$) based on Gaussian re-sampling of the observational data.
We reconstruct solely within the intercept redshift interval shared by the basis data and the data for comparison in our Bayesian analysis, considering that interpolation is more trustworthy than extrapolation.  

In Figure~\ref{fig:rec} we show the reconstructed $c/H(z)$ and $D_C(z)$ which are based on the OHD. The GP and ANN behave different on $c/H(z)$ reconstruction at both lower and higher end of the redshift range, which result to the difference of the $D_C(z)$. Based on the reconstruction, we calibrate the HIIGx Hubble diagram without assuming the flatness of the Universe and show the involved parameters in Figure~\ref{fig:ohd+hii} and their summary in Table~\ref{Table:results}.   We observe inverse correlation ($r\sim-0.49$) between $H_0$ and $\Omega_k$, which is known as the geometry degeneracy. We also observe strong inverse correlation ($r\sim-0.98$) between $\alpha$ and $\kappa$. The correlation between the cosmic parameters and the calibration coefficients should not be overlooked, especially for the correlation between $\Omega_{k0}$ and $\alpha$ that shows correlation coefficient $r\sim0.36$. As a comparison result, we also show the calibration coefficients from assuming a flat Universe and find 1.07 $\sigma$ and 1.14 $\sigma$ difference between the two set of $\alpha$ and $\kappa$ respectively.

We show the OHD calibrated HIIGx Hubble diagrams and their reconstructions in Figure~\ref{fig:mu}, companion with the calibrated JLA SNe Ia Hubble diagrams based on the reconstructed data. The calibration parameters and the jointly constrained $H_0$ are shown in Figure~\ref{fig:hii+sne} and summarised in Table~\ref{Table:results}. We observe strong correlation ($r\sim0.92$) between $H_0$ and $M^1_B$.

\begin{deluxetable*}{cccccc}
\tablecaption{Calibration Parameters of Hubble Diagrams.}\label{Table:results}
\tablehead{\colhead{HIIGx} & \colhead{$H_0$} & \colhead{$\Omega_{k0}$} & \colhead{$\alpha$} & \colhead{$\kappa$}  &  
} 
\startdata
\addlinespace
OHD+GP& $67.0\pm 2.4$ & $-0.04^{+0.29}_{-0.24}$ & $4.877^{+0.103}_{-0.098}$ & $33.51\pm 0.16$\\
OHD+ANN&$64.4^{+3.9}_{-3.4}$ & $0.17^{+0.42}_{-0.37}$ & $4.861^{+0.101}_{-0.099}$ & $33.57\pm 0.17$ \\
\addlinespace
\toprule
SNe & $H_0$ & $\alpha_x$ & $\beta$ & $M_{B}^1$ & $\Delta M$\\
\addlinespace
\hline
\addlinespace
HII+GP& $68.7^{+2.6}_{-2.5}$ & $0.1419\pm 0.0058$ & $3.071^{+0.074}_{-0.070}$ & $-18.987^{+0.073}_{-0.074}$ & $-0.071\pm 0.021$\\
HII+ANN& $65.9^{+3.0}_{-2.9}$ & $0.1422\pm 0.0058 $ & $3.068\pm 0.073$ & $-19.078^{+0.094}_{-0.093}$ & $-0.071\pm 0.021$\\
\addlinespace
\enddata
\flushleft{
\tablecomments{ Parameters are the summaries of the Figure~\ref{fig:ohd+hii} and \ref{fig:hii+sne}. $H(z)$ is in the unit of $\rm km\ s^{-1}\ Mpc^{-1}$. }}
\end{deluxetable*}

\section{Conclusion and Discussion}
\label{sec:conc}

Cosmic Chronometers (also known as OHD) can be used as distance anchors to constrain the $L-\sigma$ relation coefficients of the HII Galaxies, therefore provide us an alternative cosmological-model independent method to calibrate the HIIGx Hubble diagram. We promote this method by removing the flat Universe assumption and observe weak correlations between the cosmic parameters ($H_0$ and $\Omega_{k0}$) and the coefficients ($\alpha$ and $\kappa$). We observe approximately 1 $\sigma$ difference between the calibrations assuming or not the flatness of the Universe. This indicates the necessity of assumption-free for improving the reliability of the calibration. 

We implement a non-parametric Artificial Neural Network method for reconstructing data in the intermediate process of calibration and  obtain $\alpha = 4.861_{-0.099}^{+0.101}$ and $\kappa = 33.57 \pm 0.17$ with its reliability affirmed by highly compatible Gaussian Process involved result. We obtain Hubble constant with a value of $H_0=64.4_{-3.4}^{+3.9}~\mathrm{km~s^{-1}~Mpc^{-1}}$, which is lower but still compatible with the Planck18 measurement ($H_0=67.4\pm0.5~\mathrm{km~s^{-1}~Mpc^{-1}}$, \cite{Planck2020AA}).

We calibrate the JLA SNe Ia Hubble diagram using the calibrated HIIGx Hubble diagram as an anchor, making it possible to be free of the conventional assumption of $H_0=70~\mathrm{km~s^{-1}~Mpc^{-1}}$. The newly calibrated  JLA Hubble diagram can be used to constrain any cosmological parameters including those degenerated with the $H_0$. We obtain a jointly constrained value of $H_0=65.9_{-2.9}^{+3.0}~\mathrm{km~s^{-1}~Mpc^{-1}}$, which is more compatible with the Planck18 measurement than what we get from the HIIGx Hubble diagram.

In conclusion, our method further improves the reliability of the HIIGx Hubble diagram calibrated with OHD by removing the assumption of cosmological flatness in the method of \cite{Wu:2020} and the assumption of kernel functions in the reconstruction. This method can be used to obtain cosmological-model independent calibrations of other Hubble diagrams. 
\begin{acknowledgments}
We appreciate the referee's several useful suggestions, as well as Guo-Jian Wang's kindly instruction about the \textit{ReFANN}. We also thank Yu-Chen Wang, Yun-Long Li and Cheng-Zong Ruan for valuable discussions. This work is supported by the National Science Foundation of China(Grants Nos. 61802428, 11929301). Kang Jiao is also funded by the China Scholarship Council (CSC) from the Ministry of Education of P.R. China.
\end{acknowledgments}

\software{\textsc{GaPP} \citep{Seikel:2012},
        \textsc{ReFANN} \citep{Wang:2020j},
        \textsc{emcee} \citep{Foreman-Mackey:2013},
        \textsc{ChainConsumer} \citep{Hinton2016JOSS},
        \textsc{Numpy} \citep{Harris2020},
        \textsc{Scipy} \citep{2020SciPy-NMeth},
        \textsc{Matplotlib} \citep{Hunter2007}.
        }

\bibliography{zhangHII}{}
\bibliographystyle{apj}


\end{CJK*}
\end{document}